\documentclass[prb,twocolumn]{revtex4}
\usepackage{amssymb}
\usepackage{graphicx}
\usepackage{amsmath}

\begin{document}

\title{Raman spectroscopy study of internal dynamics in Pb(Zn$_{1/3}$Nb$_{2/3} $)O$_{3}$ and 95.55\%Pb(Zn$_{1/3}$Nb$_{2/3} $)O$_{3}$-4.5\%PbTiO$_{3}$  crystals}
\author{Oleksiy Svitelskiy}
\altaffiliation{On leave from the Institute of Semiconductor Physics, Kyiv 03028, Ukraine}
\affiliation{National High Magnetic Field Laboratory, Florida State University, Tallahassee, FL 32310, USA}
\author{Duangmanee La-Orauttapong}
\affiliation{Department of Physics-Science, Srinakharinwirot University, Bangkok 10110, Thailand}
\author{Jean Toulouse}
\affiliation{Department of Physics, Lehigh University, Bethlehem, PA 18015, USA}
\author{W.Chen}
\affiliation{Department of Chemistry, Simon Fraser University, Burnaby, BC, V5A 1S6, Canada}
\author{Z.-G.Ye}
\affiliation{Department of Chemistry, Simon Fraser University, Burnaby, BC, V5A 1S6, Canada}

\begin{abstract}
Pb(Zn$_{1/3}$Nb$_{2/3}$)O$_{3}$ is one of the simplest representatives of the lead ferroelectric relaxors. Its solid solution with PbTiO$_{3}$, at concentrations near the morphotropic phase boundary, is especially important. In this paper, we apply the tested earlier on Pb(Mg$_{1/3}$Nb$_{2/3}$)O$_{3}$ approach, to analyze the light scattering from both, nominally pure Pb(Zn$_{1/3}$Nb$_{2/3} $)O$_{3}$ crystal and a crystal containing 4.5\% of PbTiO$_{3}$, measured in a broad temperature range from 1000 to 100 K.  We propose a comprehensive picture of the temperature evolution of the lattice dynamics in these crystals and associated structural transformations.  We show, that in PZN, like in PMN, short-lived dynamic lattice distortions exist even at the highest measured temperatures. With cooling down, these distortions develop in Fm$\overline{3}$m clusters and R3m polar nanoregions that are still capable of reorientational motion. From the Burns temperature T$_{d}$, this motion becomes progressively restricted. The freezing process starts at the temperature T$^{*}$ and continues down to T$_{d0}$. The major Raman lines in all these materials behave very similarly. However, the temperature behavior of a weak line E in each case is essentially different. Its splitting can serve as a measure of the order parameter of the system.
\end{abstract}

\maketitle

\section{Introduction}

In the recent years, there is an inexhaustible interest to the investigation of the ferroelectric lead relaxors. This interest is determined, on the one hand, by their unique and important properties: dielectric, piezoelectric, electro-optic, polarizing, etc. On the other hand, researchers are attracted to the challenging task of understanding these highly complex compounds, characterized by strong compositional, chemical and structural disorder. Solving this task is of fundamental importance; it will be a large step toward understanding the disorder phenomena in general. Especially significant efforts are given to exploring internal (both, phonon and relaxational) dynamics in these crystals by neutron, X-ray and light scattering spectroscopy. However, many of the key aspects of the problem remain controversial\cite{shirane, yekey}. 

In the previous paper\cite{svit-PMN}, we have made a detailed introduction to the problem from the viewpoint of Raman spectroscopy, and shown the results of the comprehensive analysis of the Raman spectra of Pb(Mg$_{1/3}$Nb$_{2/3}$)O$_{3}$ (PMN), one of the most famous representative of the lead relaxor family. Based on the concept developed for a model relaxor material KTa$_{1-x}$Nb$_{x}$O$_{3}$ (KTN)\cite{svit-KTN},  we have presented a picture of the thermal evolution of the internal dynamics and structural transformations in PMN crystal. In crystals of cubic symmetry, first order Raman scattering may exist only due to the lattice distortions\cite{long}. We have shown that in PMN, these distortions in the dynamic form are present even at a temperature as high as 1000 K. Below the Burns temperature T$_{d}$, their dynamics becomes progressively more restricted. From the temperature T$^{*}$, the freezing process starts and continues down to the temperature of the electric field induced phase transition T$_{d0}$.

In this work, a similar analysis was applied to the light scattering spectra of Pb(Zn$_{1/3}$Nb$_{2/3}$)O$_{3}$ (PZN) crystal, which is known as the closest  \textquotedblleft relative\textquotedblright\ of the PMN in the lead relaxor family\cite{gehr}. 
Solid solutions of the lead relaxors with PbTiO$_{3}$ (PT) at concentrations close to the morphotropic phase boundary are especially important for industry\cite{gop-2}. In order to explore the influence of the PT admixture on the temperature evolution of the light scattering spectra and on the structural changes in the crystal, the results obtained on the nominally pure PZN are compared to the results obtained on a crystal containing 4.5\% of PT (PZN-4.5\%PT). Our goal is to understand, which of the properties are common to a broad class of lead relaxors, and which of them reflect their individual peculiarities. 

\begin{figure*}[t]
\includegraphics[width=1.8\columnwidth]{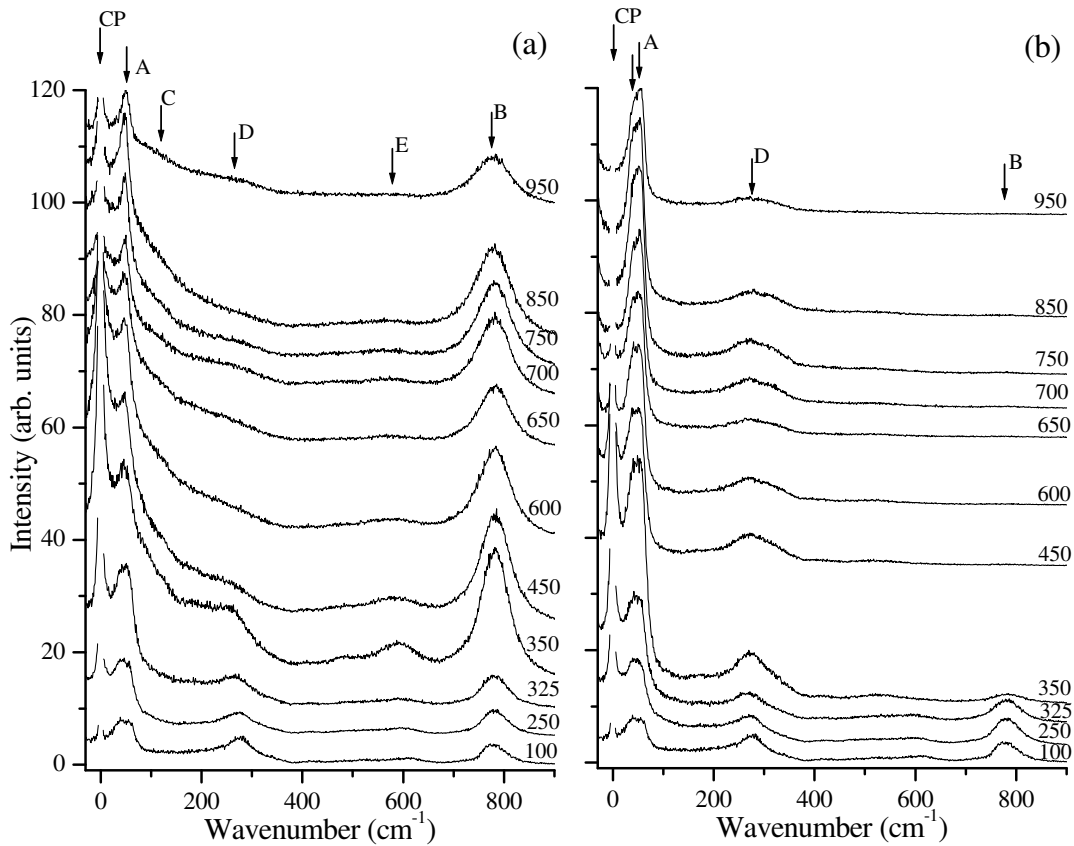}
\caption{Examples of Raman scattering spectra of PZN crystal measured upon cooling at various temperatures (see labels at the right end of the plots), in VV or $<x|zz|y>$ (a) and VH or $<x|zx|y>$ (b) geometries. }
\label{PZN}
\end{figure*}

At high temperature, PZN has a perovskite cubic structure Pm$\overline{3}$m with Zn and Nb ions interchanging on B sites. Like in PMN, its spatial composition is inhomogeneous and characterized by presence of small clusters ($\sim $2.5~nm) with occupation ratio Zn:Nb=1:1\cite{fanning-2}. Lowering temperature leads to the formation of Fm$\overline{3}$m symmetry in the 1:1 clusters and to the appearance of the polar nanoregions (PNR's), which are capable of orientational dynamics through a broad temperature range. At the Burns temperature T$_{d}\sim$750~K, the growing PNR's cause nonlinearity in the temperature dependence of the optical refractive index \cite{burnsRefraction-2}. The PNR's are responsible for the overdamping of the TO$_{1}$ phonon\cite{gehring} and for the appearance of the diffuse neutron scattering\cite{gop-PZN}. As in PMN, the development of these regions is strongly influenced by the random fields resulting from the compositional fluctuations. Consequently, a ferroelectric phase transition does not occur and the dielectric constant exhibits a broad frequency-dependent maximum at T$\sim$410~K\cite{mulvi1-2}.

Unlike PMN, whose structure exhibits an average Pm$\overline{3}$m symmetry down to the lowest temperatures, PZN (at T$\sim$385~K) has been reported to undergo a structural $non-ferroelectric$ phase transition from a cubic to a rhombohedral R3m phase \cite{lebon-2}. This phase exists only on a local scale, and the crystal does not show spontaneously formed macrodomains\cite{kamzina-2}. Application of an electric field in the $<$111$>$ direction with a threshold value of 0.7~kV/cm can induce formation of such domains and the development of a macroscopic polar phase at T$_{d0}\sim$330~K\cite{mulvi2-2}. 

PbTiO$_{3}$ (PT) is a classic soft-mode ferroelectric. Upon cooling, at T$\sim$760~K, it transforms from a cubic paraelectric phase to a tetragonal P4mm ferroelectric phase\cite{gop-2}. PT admixture to PZN causes reduction of its relaxor properties, leads to appearance of ferroelectric domains and induces a ferroelectric transition. A crossover from the relaxor to ferroelectric state is of a special interest. It occurs at PT concentration of $\sim$4-7\% \cite{gop-2}. Depending on the exact value, such a crystal transforms from cubic to rhombohedral phase either directly (as in this work) or through an intermediate ferroelectric tetragonal phase\cite{forrester}. At the same time, it still exhibits characteristic relaxor properties \cite{samara-2}. 

\begin{figure*}[t]
\includegraphics[width=1.8\columnwidth]{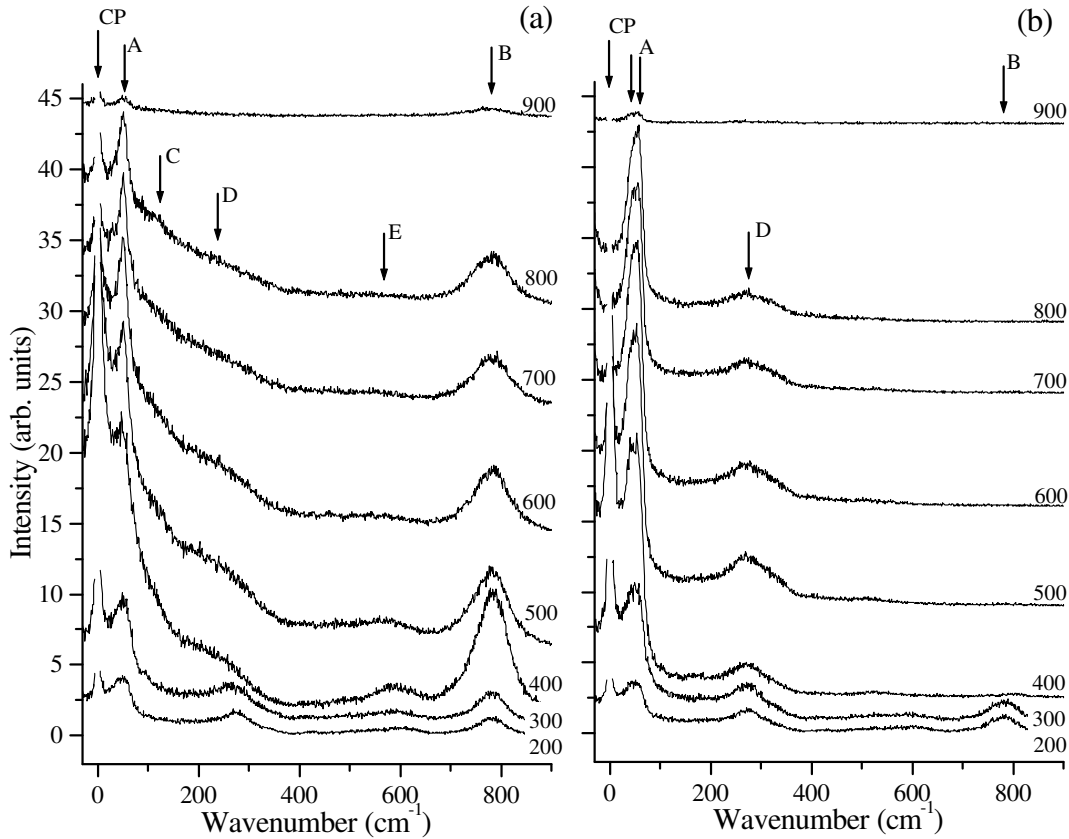}
\caption{Examples of Raman scattering spectra of PZN-4.5\%PT crystal 
measured upon cooling at various temperatures (see labels at the right end of the plots), in VV or $<x|zz|y>$ (a) and VH or $<x|zx|y>$ (b) geometries.}
\label{PZNPT}
\end{figure*}

The Raman spectra of PZN look similar to those of PMN \cite{jiang-2, lebon-2, ohwa-2}. The biggest difference occurs at low temperature, when PZN enters microdomain state of R3m symmetry with characteristic microdomain size of the order of the wavelength of light. The multiple light scattering at the boundaries of the microdomains causes a loss of polarization information, so the spectra measured in different geometrical configurations appear to be identical. The interpretation of Raman spectra in PZN, like in  PMN, is not clear (for details see Introduction in Ref. \cite{svit-PMN} and review \cite{siny-review}). Phonon assignment of particular lines, role of disorder and relaxations, and their interaction with phonon modes have yet to be understood. 

In this work, we extend our concept developed for PMN to the case of PZN-type compounds. We offer a detailed analysis of the polarized $<x$|zz|$y>$ (VV) and depolarized $<x$|zx|$y>$ (VH) Raman spectra of a single nominally pure PZN crystal and a crystal containing 4.5\% of PT. To obtain a complete picture, we made our analysis in a broad spectral (up to 1000 cm$^{-1}$) and temperature (from 100 to 1000 K) ranges. This allows us to build a comprehensive picture of the temperature evolution of the structure of PZN-type crystals.
    
\section{Experiment}

We have investigated $<$100$>$ cut single crystals. Crystal of PZN was grown by a spontaneous nucleation from high temperature solution using an optimized flux composition of Pb and B$_{2}$O$_{3}$ \cite{ye}. The PZN-4.5\%PT crystal was grown by the top-cooling solution growth technique, using PbO flux \cite{ye-1}. Both as-grown crystals were of a light-yellow color and exhibited  high optical quality. To ensure that the current results are comparable to the earlier PMN ones\cite{svit-PMN}, a special attention was paid to carry out measurements in the same experimental conditions.  The scattering was excited by focused to a 0.1~mm spot 514.5~nm light of a 300~mW Ar$^{+}$-ion laser, propagating in $\left\langle100\right\rangle $ direction. The scattered light was collected at an angle of 90$^{\circ }$ with respect to the incident beam (i.e., in $\left\langle 010\right\rangle $ direction) by a double-grating spectrometer equipped with a photomultiplier. For most of the measurements, the slits were opened to 1.7~cm$^{-1}$. Each polarization of the scattered light, $\left\langle x|zz|y\right\rangle $ (VV) and $\left\langle x|zx|y\right\rangle $ (VH), was measured separately. In order to exclude differences in sensitivity of the monochromator to different polarizations of the light, a circular polarizer was used in front of the entrance slit. For control purposes, we also took measurements without polarization analysis. To protect the photomultiplier from the strong Raleigh scattering, the spectral region from -4 to +4~cm$^{-1}$ was excluded from the scans. The data were collected in the temperature range from 1000 to 100~K. The cooling rate was 0.5-1~K/min. Every 50~K the temperature was stabilized and the spectrum recorded.  For clarity, we describe the observed phenomena from high to low temperatures, following the same order as in measurements (unless otherwise stated).

Figure \ref{PZN} demonstrates examples of PZN Raman spectra measured in VV [panel (a)] and VH [panel (b)] geometries at different temperatures. These spectra are consistent with the previously reported\cite{iwata, ohwa}. Like in PMN, at high temperature a typical spectrum consists of two strong lines centered approximately at 45~cm$^{-1}$ and 780~cm$^{-1}$ (labeled A and B), three broad bands (C, D, E) and of a central peak (CP). Line A exhibits a fine structure, which is seen explicitly from a comparison of the spectra measured in VV and VH polarizations. Lowering temperature modifies shapes of the broad bands C, D and E. However, these modifications are not as strong as in the case of PMN (compare Fig. \ref{PZN} and Fig. 2 from Ref. \cite{svit-PMN}).

Our results confirm\cite{iwata-cub} existence of scattering even at a temperature of 950~K (Fig. \ref{PZN}). Cooling the crystal, the intensity of the scattering, first grows (compare plots for 950, 850 and 750~K); starting from the Burns temperature T$_{d}\sim$750~K it decreases (plots for 700 and 650~K), and then increases again. At the temperature of 600~K, the intensity is back to its value at 750~K and continues to grow till $\sim$350~K. Below this temperature, the crystal enters a microdomain state. Multiple scattering on the domain walls causes general decrease of intensity and loss of polarization information (VV and VH became approximately equal). This multiple scattering is significant when average domain size is comparable to the wavelength of light. In our case, the temperature of its appearance closely correlates with the prediction from the diffused neutron scattering experiments \cite{gop-PZN}. The intensity decrease observed below this transition is determined primarely by the Bose population factor: 

\begin{equation}
F(f,T)=\left\{ 
\begin{array}{c}
(e^{hf/kT}-1)^{-1}+1,\text{\ Stokes part} \\ 
(e^{hf/kT}-1)^{-1},\text{\ anti-Stokes part}%
\end{array}%
\text{ }\right. \text{, }  \label{boze}
\end{equation}%

\noindent and the Raman scattering demonstrates a characteristic first-order behavior. Similar shapes of the low- and high- temperature spectra show that even at 950~K the scattering exhibits a first-order character. In the perfectly cubic crystal, first-order scattering is prohibited. Therefore, one should assume the presence of distortions in the form of lower symmetry clusters, like Fm$\overline{3}$m and R3m. At high temperature, these clusters are dynamic, having short lifetimes. Lowering the temperature, they become progressively longer lived or quasidynamic. In PZN, the observed decrease in Raman intensity around T$_{d}$ is not as strong as in PMN (see Fig. 2 in Ref. \cite{svit-PMN}). We associate it with the formation of Fm$\overline{3}$m clusters in their evolution from a dynamic to a static form. With further cooling, the PNR's grow; their dynamics slows down. This explains the changes in the shapes of a number of peaks. Below T$_{d0}\sim$330~K, which is close to the temperature of the electric field-induced ferroelectric phase transition, the shape of the spectra is relatively stable. This means that the PNR's became static and the process of their formation is over.

Figure \ref{PZNPT} demonstrates how admixture of 4.5\% of PT influences Raman scattering of PZN. Panel (a) shows spectrum measured in VV and panel (b) in VH geometries at different temperatures. These spectra are consistent with those measured earlier\cite{gupta, iwata-cub}. Their shapes look very similar to the shapes of nominally pure PZN (Fig. \ref{PZN}). However, even such a relatively small amount of PT causes significant modifications. First, a high temperature weakening of the scattered intensity (compare the data measured at 900 and 800~K). In our case, this intensity decrease occurred at $\sim$200 degrees lower temperature then in the earlier study\cite{iwata-cub}. It has to be attributed to a disappearance of lattice distortions and establishing a  \textquotedblleft pure\textquotedblright\ Pm$\overline{3}$m cubic symmetry in the lattice, which does not allow first-order scattering. This result is rather unexpected, since PT admixture facilitates appearance of PNR's causing lattice distortions. To understand it, one has to assume a weakening (in comparison to the thermal energy $kT$) of the interatomic bonds in the PT-doped PZN crystal at the high temperatures. Second change that is seen at a glance, is the absence of the described above decrease of intensity related to the formation of Fm$\overline{3}$m clusters. This suggests hindering the development of the Fm$\overline{3}$m clusters under the influence of PT, so either the size of the clusters or their density is insufficient to worsen the optical quality of the sample. A further analysis allows us to reveal a number of other interesting properties of these compounds.

\begin{figure}[th]
\includegraphics[width=1.1\columnwidth]{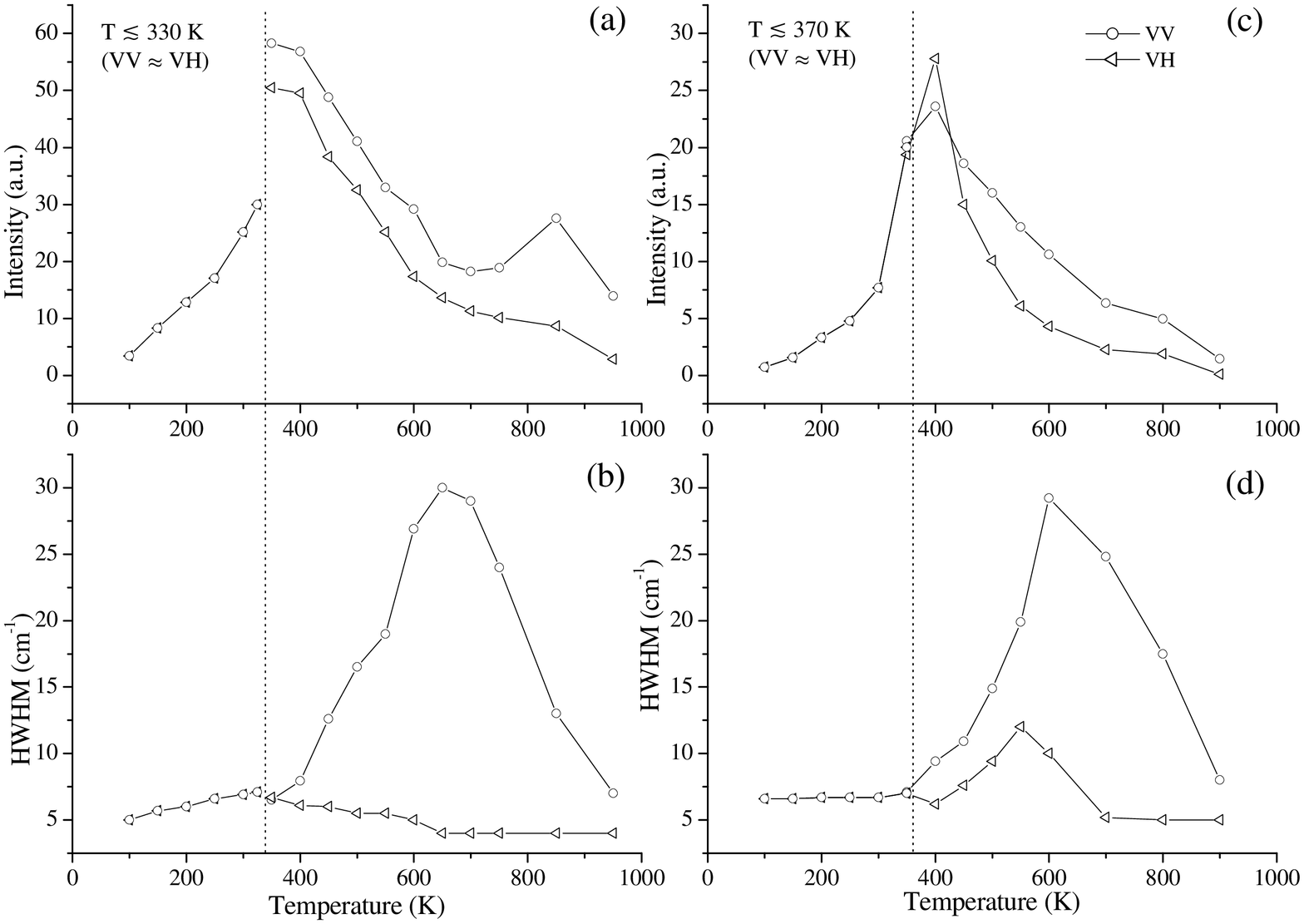}
\caption{Temperature evolution of the parameters of the central peak (Lorenzian approximation) in PZN [left panel - (a) and (b)] and PZN-4.5\%PT [right panel - (c) and (d)] crystals. Intensities are presented in (a) and (c), half widths at half maxima are shown in (b) and (d).}
\label{CP}
\end{figure}

\section{Multiple-peak decomposition and most interesting results of it}

The measured data was analyzed by a multiple peak decomposition, following the same procedure as in the PMN work\cite{svit-PMN}. The central peak (CP) was assumed to have a shape of Lorentzian, other peaks were described as damped harmonic oscillators multiplied by the Bose factor (\ref{boze}):

\begin{equation}
\Phi _{i}\sim \frac{\Gamma _{i}f_{0i}^{2}f}{(f^{2}-f_{0i}^{2})^{2}+\Gamma
_{i}^{2}f_{0i}^{2}}F(f,T)\text{ .}
\end{equation}

\noindent Here $\Gamma _{i}$ and $f_{0i}$ are the damping constant and the mode frequency. Consider the most interesting results of this analysis.

\subsection{Central peak}

In Raman spectrum, a CP can appear either due to relaxations\cite{siny-cp, sokol-1}, or due to the zone-center softening of the TO phonon\cite{fontana-1}, or due to interaction of these two processes\cite{vugm}. Because the zone center TO phonon is overdamped\cite{gehring}, we may suppose that our central peak originates solely from relaxations.  When the relaxations are fast\cite{MichelA}, the CP is non-intense and broad. Their slowing down causes the growth and narrowing of the CP. Such an approach has proven to be profitable for investigations of internal motion not only in relaxor ferroelectrics\cite{siny-cp, sokol-1}, but also in other materials with relaxational dynamics\cite{RoweA, LutyA}. The temperature evolution of the parameters of the Lorentzian approximation of the CP in the investigated PZN (left panel) and PZN-4.5\%PT (right panel) crystals is presented in Fig. \ref{CP}. Panels (a) and (c) show intensities, (b) and (d) demonstrate half widths at half maxima of the peak. 

\begin{figure}[th]
\includegraphics[width=1.1\columnwidth]{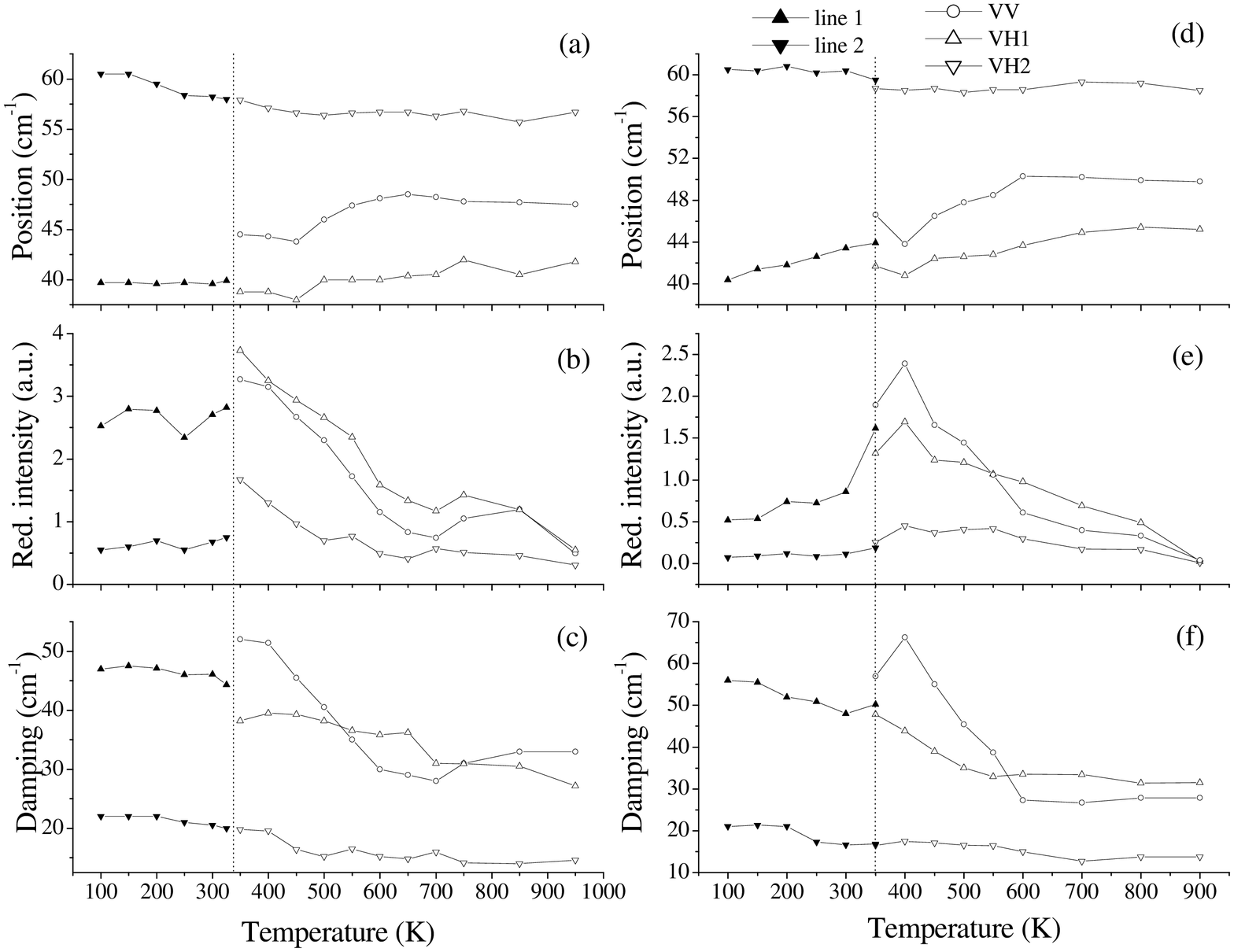}
\caption{Temperature evolution of the components of the peak A. Left hand side shows their positions (a), reduced intensities (b) and damping constants (c) for PZN crystal; right hand side (d, e and f) shows these parameters for PZN-4.5\%PT crystal. The dashed lines separate dynamic and static phases. To its left, the polarization information was lost.}
\label{peakA}
\end{figure}

The temperature behavior of the CP in PZN is similar to that in PMN\cite{svit-PMN} and KTN\cite{svit-KTN} relaxor crystals.  We have shown\cite{svit-PMN, svit-KTN, grace} that the evolution of the CP in those materials can be explained by a model involving the relaxational motion of the polar nanoregions and its progressive restriction with decreasing temperature: from possibility of reorientational motion amongst eight equivalent $<$111$>$ directions at the high temperature, down to confinement to a single allowed direction at the low-temperature. The similarity of the CP behavior in these different materials, suggests that the polar clusters in PZN and PZN-4.5\%PT as well as in PMN and KTN develop through a similar sequence of stages, although not necessarily causing the appearance of the long-range ferroelectric order. 

Starting from high-temperature, the first important feature is the local VV intensity maximum at the temperature $\sim$850~K (Fig. \ref{CP}(a)), accompanied by a relatively weak VH component. From comparison with PMN and KTN, where this feature is much stronger, we attribute this phenomenon to the symmetric 180$^\mathrm{o}$ reorientational motion of ions. Further lowering the temperature, a cessation of this motion causes some decrease in the VV intensity of the CP, reaching a minimum in the neighborhood of the Burns temperature T$_{d}\sim$750~K. Like in PMN, the prohibition of 180$^\mathrm{o}$ reorientations introduces anisotropy into the lattice causing the distinguishability between B sites occupied by different ions, Zn and Nb, and the appearance of a superstructure with, on average, Fm$\overline{3}$m symmetry in the 1:1 ordered areas. It also puts restrictions on the reorientational motion of the dynamical R3m polar nanoregions. Now, they can reorient only amongst four neighboring $<$111$>$ directions, introducing local time-averaged tetragonal distortions. These processes are accompanied by the appearance of large (of the order of wavelength of light) quasidynamic fluctuations causing some worsening of the optical quality of the sample. With further cooling, the optical quality of the crystal improves. From $\sim$650~K, the four-site reorientational motion of the PNR's is slowing down, leading to the narrowing of the VV component of the CP (Fig. \ref{CP}(b)), the increase of its VV and VH intensities (Fig. \ref{CP}(a)) and the simultaneous broadening of the VH component (Fig. \ref{CP}(b)). These indicate the appearance of the new restrictions on the ionic motion in the crystal. Analogy with KTN and PMN suggests that the motion of R3m clusters gradually becomes restricted to two neighboring $<$111$>$ orientations, averaging to monoclinic distortions. However, in PZN this process develops much more gradually, without a definite temperature separating the stages with predominantly four- and two- site motion. Comparing with the temperature behavior of the CP in PMN and with neutron scattering results, we would approximate it as T$^{*}\sim$500~K. Moreover, with further cooling a gradual freezing of the reorientational motion begins, leading to the appearance and growth of the static clusters with R3m symmetry. The intensities of both, VV and VH, components of the CP reach their maxima at T$\sim$350~K. At T$\sim$330~K, when the characteristic sizes of the R3m regions become comparable to the wavelength of light (also see Fig. 3 in Ref. \cite{gop-PZN}), the worsening of the optical quality of the sample causes loss of the polarization information and general decrease in the scattered intensity. Note, that this temperature coincides with the temperature of the electric field induced phase transition T$_{d0}$. Thus, in PZN, like in PMN, the temperature T$_{d0}$ marks the end in the structural transformations sequence. Below T$_{d0}\sim$330~K, the CP is narrow and its intensity continues to decrease.

Despite the central peak of PZN-4.5\%PT crystal looks very similar to the one of PZN, admixture of PT introduces significant changes in its temperature behavior (Fig. \ref{CP}(c and d)). The local high-temperature maximum in the VV intensity of the CP, which was clearly present in KTN (Fig. 3 in Ref. \cite{svit-KTN}), PMN (Fig. 4 in Ref. \cite{svit-PMN}), and PZN (Fig. \ref{CP}(a)), in PZN-4.5\%PT is absent. By analogy with other crystals, this local maximum should be expected at T$\sim$800~K. However, at this temperature only a monotonic growth of the VV intensity with just a little kink (if any) is observed (Fig. \ref{CP}(c)). This confirms the expressed above suggestion that addition of PT strengthens restrictions on the reorientational motion, suppressing 180$^\mathrm{o}$ reorientations. Thus, in PZN-4.5\%PT the PNR's are restricted to a four-cite reorientational motion from the moment of their nucleation. Further cooling leads to increase in the VV and VH intensities (Fig. \ref{CP}(c)) and broadening of the central peak (Fig. \ref{CP}(d)). Like in PMN (Fig. 4 in Ref. \cite{svit-PMN}), the half-widths of the VV and VH components of the peak have strong maxima occurring at two different temperatures ($\sim$600 and 550~K respectively). This indicates that in PZN-4.5\%PT the change in the reorientational motion of the PNR's from four- to two- direction character occurs more sharply then in nominally pure crystal. It can be associated with a particular temperature T$^{*}\sim$550~K (which is close to the estimate from the neutron scattering studies\cite{gop-PZNPT}) of an "underlying" structural transformation, below which the PNR's are restricted to reorient only between two allowed $<$111$>$ directions. Then, their motion starts to freeze. The size of the clusters is growing and, at T$_{d0}\sim$370~K their reduce the optical quality of the sample leading to decrease in Raman intensity and loss of the polarization information. Below this temperature, the CP is narrow, and its intensity decreases.

\begin{figure}[th]
\includegraphics[width=1.1\columnwidth]{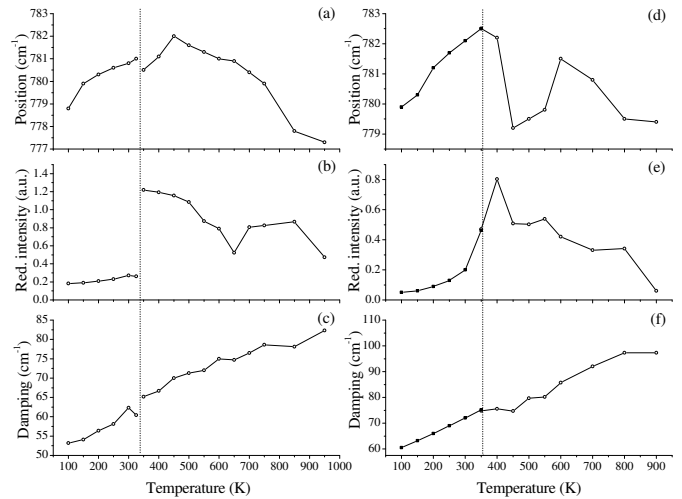}
\caption{Temperature evolution of the peak B. Left-hand side shows its position (a), reduced intensity (b) and damping constant (c) for PZN crystal; right-hand side (d, e and f) shows these parameters for PZN-4.5\%PT crystal. The dashed lines separate dynamic and static phases. Below it the polarization information is lost.}
\label{peakB}
\end{figure}

\subsection{The major peaks A and B}

Figure \ref{peakA} presents the temperature evolution of the fitting parameters of the peak A. The left panel (a-c) shows their values for the nominally pure PZN. The right panel (d-f) demonstrates parameters for the peak A in PZN-4.5\%PT. Top plots (a and d) show positions, middle (b and e) reduced intensities, bottom (c and f) present the values of damping constants. The dashed lines separate the dynamic and static regions. For both compositions, at temperatures above the dashed line, this peak has a triplet structure, containing one component in VV (circles) and two components in VH (light up and down triangles) geometry. Below the dashed lines, the polarization information was lost and the peak appeared as a doublet (dark up and down triangles). 

Except for the loss of polarization in the static phase, the components of the peak A of the nominally pure PZN exhibit a strikingly similar temperature behavior to those of PMN (compare to Fig. 5 (a-c) in \cite{svit-PMN}). The detailed analysis\cite{svit-PMN} of this peak in PMN allowed to attribute its the origin to the interaction of the zone center TO and zone-boundary TA phonons with different polarizations. Such an interaction is possible if it is mediated by the disordered lattice distortions, which explains the increase of damping with lowering temperature. The mediation is also important in explaining the presence of a rather strong scattering signal in the VH geometry. This idea has further development in the first-principle calculations, that are currently in progress\cite{pros}. The similarities of the parameters of this peak in PMN and PZN emphasize the common origin of the peak A in these two relaxor materials. Admixture of 4.5\% of PT (Fig. \ref{CP}(d-f)) does not cause any dramatic changes in the temperature trends of the components of the peak A. Primarily, its influence is limited to sharpening of the features related to the formation of the static phase (like those around 400~K). 

\begin{figure}[th]
\includegraphics[width=1\columnwidth]{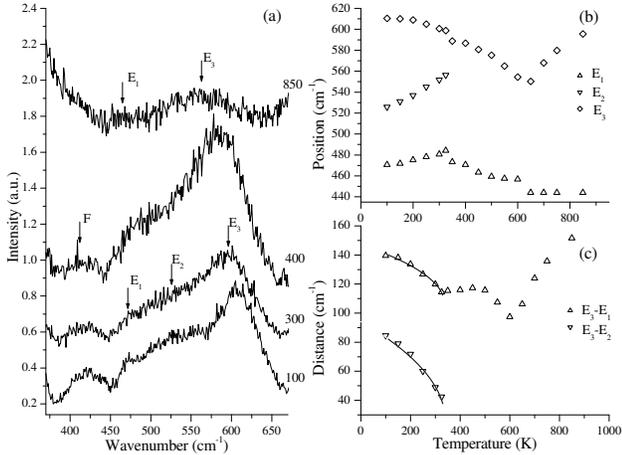}
\caption{Peak E in PZN crystal: examples of the spectra measured at several temperatures (shown on the labels at the right end of the plots); temperature evolution of the positions of the peak components  (b). Distances between these components (c) demonstrate that splitting of this peak follows critical dependence law. Solid lines represent (T$_{d0}$-T)$^{1/2}$ fit.}
\label{EPZN}
\end{figure}

Figure \ref{peakB} demonstrates temperature behavior of the line B located at the high-frequency end of the spectrum. The left panel (a-c) shows its parameters for PZN; the right panel (d-f) for PZN-4.5\%PT. Like in the case of peak A, the top plots (a and d) show positions, the middle plots (b and e) show reduced intensities, and the bottom plots (c and f) show damping coefficients. For both materials, in the dynamic region, this line has very strong intensity in the VV geometry, but weak in the VH one (i.e. characterized by A$_{1g}$ symmetry). Below transition to the static phase, VV and VH intensities are equal (also see Figs. \ref{PZN}, \ref{PZNPT}). Comparison of the left panel of Figure \ref{peakB} with the right panel of Figure 6 in Ref. \cite{svit-PMN} shows that except for the loss of polarization at the transition to the static state, parameters of the peak B in PZN exhibit the same temperature trends as in PMN; following the formation of dynamic polar clusters at T$_{d}$, their slowing down, and gradual development of the static order from T$^{*}$ to T$_{d0}$. Addition of 4.5\% of PT leads to appearance of a new feature - lowering energy (softening) of this line [Fig. \ref{peakB}(d)] around 500~K. As we explained above, this temperature is associated with the growing restrictions on the dynamics of PNR's. It is widely accepted that PNR's are related to the off-centered Nb ions. Therefore, the softening of the peak B is a direct confirmation of the suggestion that this line indeed originates from the oscillation in the Mg-O-Nb bond\cite{husson}.

\begin{figure}[th]
\includegraphics[width=1\columnwidth]{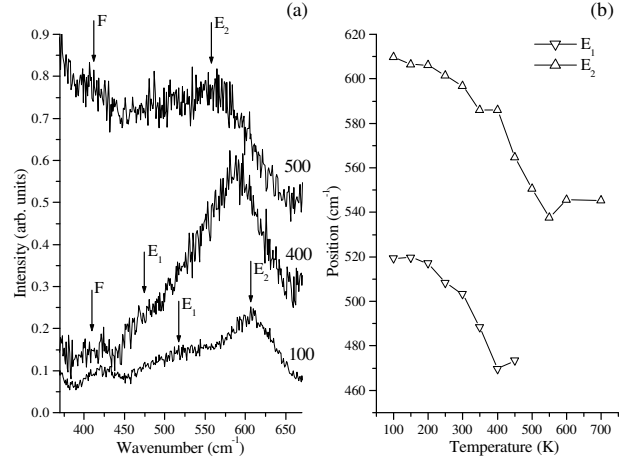}
\caption{Peak E in PZN-4.5\%PT crystal: (a) examples of the spectra measured at several temperatures (shown on the labels at the right end of the plots); (b) temperature evolution of the positions of the peak components.}
\label{EPZNPT}
\end{figure}

\subsection{The smaller peaks}

Between peaks A and B there are several other peaks, labeled C, D and E (Figs. \ref{PZN}, \ref{PZNPT}). At high temperature their intensities are very weak, and they look like merging broad bands. Lowering the temperature, the peaks become sharper and some of them show fine structure. In general, this process is similar to the one observed in PMN (Figs. 1 and 2 in Ref. \cite{svit-PMN}). However, in PZN and PZN-4.5\%PT it is not as strong and evident. Especially interesting to consider it on the example of line E. In PMN, at high temperature line E was broad and weak; with cooling the crystal, its intensity was growing; at T$<$T$^{*}$ the line became a doublet, and the temperature dependence of the distance between components was following a characteristic order parameter behavior (Fig. 6(a) in Ref.\cite{svit-PMN}), proving existence of a latent phase transformation. 

Figure \ref{EPZN}(a) shows examples of the spectra zoomed in on the line E in PZN (region between 400 and 650 cm$^{-1}$). Unlike the case of PMN\cite{svit-PMN}, in PZN line E consists of two components, E$_{1}$ and E$_{3}$, even at as high temperature as 850~K. Lowering the temperature, their intensities are growing. Starting from T$_{d0}$, one more component E$_{2}$ appears. The temperature dependences of the positions of these three components are shown in Fig. \ref{EPZN}(b). The distances between them are shown in Fig. \ref{EPZN}(c). Fig. \ref{EPZN} demonstrates that line E is very sensitive to the processes developing in the crystal while it is cooled down. The sharp feature, located around 600~K, is associated with appearance of restrictions on reorientational motion of PNR's from four- to two- directional reorientations. Below T$_{d0}\sim$330~K, the distances E$_{3}$-E$_{1}$ and E$_{3}$-E$_{2}$ both follow a (T$_{d0}$-T)$^{-1/2}$ law. This suggests that, in PZN, line E can also be an indicator of an order parameter related to appearance of the static phase.

The behavior of line E in PZN-4.5\%PT crystal is presented in Fig. \ref{EPZNPT}. Panel (a) shows examples of the spectra. At high temperature, line E is presented by a single broad component (labeled E$_{2}$). Lowering the temperature, one more component of the line becomes apparent (E$_{1}$). Unlike PZN, in PZN-4.5\%PT the third component of line E does not develop. Figure \ref{EPZNPT}(b) shows temperature dependences of the frequencies of line E components. Except for the feature around 400~K (near the ferroelectric phase transition), frequencies of the both components follow the same law while temperature decreases. Full understanding of the behavior of this line requires further progress in theoretical work.

\section{Conclusion}

A detailed multiple-peak fit analysis of the Raman spectra measured on PZN and PZN-4.5\%PT crystals in a broad temperature range
(from 1000 to 100~K), allowed us to develop the picture of the structural transformations in these crystals in the light of the earlier studies of PMN\cite{svit-PMN}. Our data show that the structural evolution of the lattice in all three compounds occurs in similar characteristic stages. Dynamical lattice distortions are present even at the highest measured temperatures. These distortions are responsible for the presence of first-order Raman scattering. At the high-temperature, the scattering is quite strong in PZN and PMN, but rather weak in PZN-4.5\%PT. Lowering the temperature, the dynamics becomes progressively more correlated. At the Burns temperature T$_{d}$, the 180$^\mathrm{o}$ reorientations become prohibited, causing appearance of the polar nanoregions and spatially averaged Fm$\overline{3}$m symmetry in the 1:1 ordered areas. The scattering intensity increases, especially in PZN-4.5\%PT so it becomes comparable to the one in PZN. With further cooling, the motion of the quasi-dynamic PNR's becomes more restricted. From the temperature T$^{*}$, they can reorient only between two possible directions, and the quasi-static ordering begins. By the temperature T$_{d0}$ the ordering process is mainly finished, resulting either in a disordered (in PMN and PZN) or in a ferroelectrically ordered (in PZN-4.5\%PT) rhombohedral phase. The shapes and the temperature behavior of the central peaks and the major phonon lines in the Raman spectra of these three materials are very similar, which emphasizes their common origin determined by processes involving phonons with different wavevectors interacting with various kinds of disorder. A rather weak phonon line E is the most sensitive to the composition of the particular crystal. In all three cases it exhibited distinctly different behavior, serving as a measure of the order parameter of the system. 

\section{Acknowledgement}

Authors are very grateful to S. Prosandeev, G. Yong, V.Dierolf, A. Souslov, E. Venger, and G. Semenova for stimulating discussions.


\bigskip

\begin{thebibliography}{99}

\bibitem{shirane} G.Shirane, G.Xu, P.M.Gehring, Conference Proc. for the NATO Advanced Research Workshop on the Disordered Ferroelectrics, Kiev, May 2003.

\bibitem{yekey} Z.G.Ye, Key Eng. Mater. \textbf{155-156}, 81 (1998).

\bibitem{svit-PMN} O.Svitelskiy, J.Toulouse, G.Yong, Z.-G.Ye, Phys. Rev. B \textbf{68}, 104107 (2003).

\bibitem{svit-KTN} O.Svitelskiy and J.Toulouse, J. Phys. Chem. Solids \textbf{64}, 665 (2003).

\bibitem{long} D.A. Long, Raman Spectroscopy, McGraw-Hill, New York, 1977.

\bibitem{gehr} P.M.Gehring, W.Chen, Z.-G.Ye, G.Shirane, arXiv:cond-mat/0304289 v1, 11 Apr 2003.

\bibitem{gop-2}  D.La-Orauttapong, B.Noheda, Z.-G.Ye, P.M.Gehring, J.Toulouse, D.E.Cox, G.Shirane, Phys. Rev. B \textbf{65}, 144101 (2002).

\bibitem{fanning-2}  D.M.Fanning, I.K.Robinson, X.Lu, D.A.Payne, J. Phys. Chem. Solids \textbf{61}, 209 (2000).

\bibitem{burnsRefraction-2} G.Burns, F.H.Dacol, Solid St. Commun. \textbf{48} (1983) 853 and Phys. Rev. B. \textbf{28}, 2527 (1983).

\bibitem{gehring} P.M.Gehring, S.-E.Park, G.Shirane, Phys. Rev. B \textbf{63}, 224109 (2001).

\bibitem{gop-PZN} D.La-Orauttapong, J.Toulouse, J.L.Robertson, and Z.-G.Ye, Phys. Rev. B \textbf{64}, 212101 (2001). 

\bibitem{mulvi1-2}  M.L.Mulvihill, S.E.Park, G.Risch, Z.Li, K.Uchino, T.Shrout, Jpn. J. Appl. Phys. \textbf{35}, 3984 (1996).

\bibitem{lebon-2}  A.Lebon, M.El Marssi, R.Farhi, H.Dammak, G.Calvarin, J.Appl. Phys. \textbf{89}, 3947 (2001).

\bibitem{kamzina-2}  L.S.Kamzina, N.N.Krainik, L.M.Sapozhnikova, S.V.Ivanova, Sov.Phys. Solid State \textbf{33}, 1169 (1991).

\bibitem{mulvi2-2}  M.L.Mulvihill, L.E.Cross, W.Cao, K.Uchino, J.Am. Ceram. Soc. \textbf{80}, 1462 (1997).

\bibitem{forrester} J.S.Forrester, R.O.Piltz, E.H.Kisi, G.J.McIntyre, J. Phys.: Condens. Matter \textbf{13}, L825 (2001).

\bibitem{samara-2} G.A.Samara, E.L.Venturini, V.H.Schmidt, Phys. Rev. B \textbf{63}, 184104 (2001).

\bibitem{jiang-2} F.Jiang, S.Kojima, Jpn. J. Appl. Phys. \textbf{38}, 5128 (1999).

\bibitem{ohwa-2} H.Ohwa, M.Iwata, N.Yasuda, Y.Ishibashi, Ferroelectrics \textbf{229}, 147 (1999); \textbf{218}, 53 (1998).

\bibitem{siny-review} I.G.Siny, S.G.Lushnikov, R.S.Katiar, V.H.Schmidt, Ferroelectrics \textbf{226}, 191 (1999).

\bibitem{ye} L.Zhang, M.Dong, and Z.-G.Ye, Mater. Sci. Eng. B \textbf{78}, 96 (2000).

\bibitem{ye-1} W.Chen and Z.-G.Ye, J.Mater. Sci. \textbf{36}, 4393 (2001).

\bibitem{iwata}  M.Iwata, H.Hoshino, H.Orihara, H.Ohwa, N.Yasuda, Y.Ishibashi, Jpn. J. Appl. Phys \textbf{39}, 5691 (2000) ;  M.Iwata, N.Tomisato, H.Orihara, N.Arai, N.Tanaka, H.Ohwa, N.Yasuda, Y.Ishibashi, Jpn. J. Appl. Phys \textbf{40}, 5819 (2001).

\bibitem{ohwa} H.Ohwa, M.Iwata, N.Yasuda, Y.Ishibashi,  Jpn. J. Appl. Phys \textbf{37}, 5410 (1998).

\bibitem{iwata-cub} M.Iwata, N.Tomisato, H.Orihara, H.Ohwa, N.Yasuda, Y.Ishibashi, Ferroelectrics \textbf{261}, 83 (2001)

\bibitem{gupta} S.Gupta, R.S.Katiyar, R.Guo, A.S.Bhalla, Journal of Raman Spectroscopy \textbf{31}, 921 (2000).

\bibitem{siny-cp} I.G.Siny, S.G.Lushnikov, R.S.Katiyar, E.A.Rogacheva,
Phys.Rev.B. \textbf{56}, 7962 (1997).

\bibitem{sokol-1} J.P.Sokoloff, L.L.Chase, D.Rytz, Phys. Rev. B, \textbf{38}, 597 (1988).

\bibitem{fontana-1} M.D.Fontana, G.E.Kugel, C.Carabatos-Nedelec, Phys. Rev. B, \textbf{40}, 786 (1989).

\bibitem{vugm} B.E.Vugmeister, Y.Yacoby, J.Toulouse, H.Rabitz, Phys. Rev. B, \textbf{59}, 8602 (1999).

\bibitem{MichelA} K.H.Michel, J.Naudts, B.De Raedt. Phys. Rev. B, \textbf{18}, 648 (1978).

\bibitem{RoweA} J.M.Rowe, J.J.Rush, S.Susman, Phys.Rev. B \textbf{28}, 3506
(1983).

\bibitem{LutyA} F.Luty, in \textit{Defects in Insulating Crystals}, edited
by V.M.Tuchkevich and K.K.Shvarts (Zinatne Publishing House, Riga, 1981)
p.70.

\bibitem{grace} G.Yong, J.Toulouse, R.Erwin, S.M.Shapiro and B.Hennion, Phys. Rev. B \textbf{62}, 14736 (2000).

\bibitem{gop-PZNPT} D.La-Orauttapong, J.Toulouse, Z.-G.Ye, W.Chen, R.Erwin, and J.L.Robertson, Phys. Rev. B \textbf{67}, 134110 (2003).

\bibitem{pros} S.Prosandeev, CAN and PMN simulation papers, work in progress

\bibitem{husson} E.Husson, L.Abello, A.Morell, Mat. Res. Bull., \textbf{25},
539 (1990).

\end{thebibliography}
\end{document}